\documentclass[reprint,amsmath,amssymb,aps,superscriptaddress,twocolumn,10pt]{revtex4-1}
\usepackage[colorlinks,linkcolor=blue,citecolor=blue,urlcolor=blue]{hyperref}
\usepackage{graphicx}
\usepackage{subfigure}
\usepackage{dcolumn}
\usepackage{bm}
\usepackage{hyperref}
\allowdisplaybreaks
\newcommand{\be}{\begin{equation}}
\newcommand{\ee}{\end{equation}}
\newcommand{\bea}{\begin{eqnarray}}
\newcommand{\eea}{\end{eqnarray}}
\newcommand{\nn}{\nonumber}

\newcommand{\TRC}{TianQin Research Center for Gravitational Physics and School of Physics and Astronomy, Sun Yat-sen University (Zhuhai Campus), 2 Daxue Rd., Zhuhai 519082, P. R. China.}
\newcommand{\HUST}{MOE Key Laboratory of Fundamental Physical Quantities Measurement \& Hubei Key Laboratory of Gravitation and Quantum Physics, PGMF and School of Physics, Huazhong University of Science and Technology, Wuhan 430074, P. R. China.}

\begin{document}

\title{Constraining modified gravity with ringdown signals: an explicit example}

\author{Jiahui Bao}
\affiliation{\TRC}

\author{Changfu Shi}
\affiliation{\TRC}
\affiliation{\HUST}

\author{Haitian Wang}

\author{Jian-dong Zhang}
\email{Emial:zhangjd9@mail.sysu.edu.cn}

\author{Yiming Hu}

\affiliation{\TRC}

\author{Jianwei Mei}
\email{Emial:meijw@sysu.edu.cn}

\author{Jun Luo}
\affiliation{\TRC}
\affiliation{\HUST}

\date{\today}

\begin{abstract}
An explicit example is found showing how a modified theory of gravity can be constrained with the ringdown signals from merger of binary black holes. This has been made possible by the fact that the modified gravitational theory considered in this work has an exact rotating black hole solution and that the corresponding quasi-normal modes can be calculated. With these, we obtain the possible constraint that can be placed on the parameter describing the deviation of this particular alternative theory from general relativity by using the detection of the ringdown signals from binary black holes's merger with future space-based gravitational wave detectors.
\end{abstract}

\maketitle

\section{\label{sec:level1}Introduction}

General relativity (GR) has enjoyed much success in passing all experimental tests to date \cite{Will:2005va}. However, theoretical problems related to the black hole singularity and black hole information and observation evidence on dark matter and dark energy all indicate that GR may not be the final theory of gravity. Numerous modified theories of gravity have been proposed to study possible extensions to GR \cite{Clifton:2011jh, Berti:2015itd}.

The first detection of gravitational wave (GW) by LIGO \cite{Abbott:2016blz} has made new tests of GR possible. Quasi-normal modes (QNMs)  \cite{Berti:2009kk, Konoplya:2011qq}, which encode all information in the ringdown stage of a compact binary merger, is especially important for the purpose \cite{Berti:2018vdi}. According to the no-hair theorem which states that all black holes in nature are Kerr black holes, the QNMs constituting the ringdown signal from a binary black hole merger are completely determined by the mass and spin of the remnant black hole. With the detection of at least two QNMs, it is possible to test the no-hair theorem by checking if the frequencies and the damping times are the same as those predicted by GR \cite{Berti:2005ys, Dreyer:2003bv}. A failure of the no-hair theorem may indicate that either GR is not the correct gravitational theory or GR is correct but the remnant is not described by the Kerr metric.

In this work, we are interested in testing a specific modified theory of gravity with the ringdown signal emitted by a binary black hole's merger. For this purpose, we will blame all possible failure of the no-hair theorem on the difference between the modified gravity theory and GR.

There have been work on constraining modified theories of gravity or no-hair theorem with GW observations.
However, so far the constraints are mostly given in terms of phenomenological parameters whose relation to each modified theory of gravity is not known explicitly \cite{Shi:2019hqa, Gossan:2011ha, Meidam:2014jpa}. In order to constrain a specific modified theory of gravity with ringdown signal, one has to overcome at least two obstacles:
\begin{itemize}
\item The first is to have reliable information on at least two QNMs through GW detection. GW150914 is the first and by far the strongest GW signal detected, with a total signal to noise ratio (SNR) reaching 24. But the SNR for the ringdown stage is only about 7, so it is extremely difficult to extract the frequencies and damping times for the subdominant mode. For future space-based GW detectors, such as LISA \cite{Audley:2017drz} and TianQin \cite{Luo:2015ght},  and the next generation ground-based detectors, such as Einstein Telescope \cite{Punturo:2010zz} and Cosmic Explorer \cite{Evans:2016mbw}, high enough SNR is possible.
\item The second is to calculate the QNMs for rotating black holes in the modified theory of gravity under study. The most promising source for testing the no-hair theorem with QNMs is the merger of massive black holes, of which the remnant black hole is usually a rotating one. But rotating black hole solutions in modified theories of gravity are difficult to find and it is also difficult to calculate the corresponding QNMs. For example, QNMs have only been studied for non-rotating or slowly rotating black holes in very few alternative theories, such as the dynamical Chern-Simons gravity \cite{Cardoso:2009pk,Molina:2010fb}, Einstein-dilaton-Gauss-Bonnet gravity \cite{Blazquez-Salcedo:2016enn, Blazquez-Salcedo:2017txk} and Horndeski gravity \cite{Tattersall:2018nve}.
\end{itemize}

Due to these difficulties, an example has been lacking where the ringdown signals from binary black holes' merger are used to constrain a specific  modified theory of gravity.

We find that in the Scalar-Tensor-Vector Gravity theory (STVG) \cite{Moffat:2005si}, a rotating black hole solution is known and the dependence of the corresponding QNMs on the key STVG parameter can also be obtained by simple means. As such, one can place explicit constraint on STVG with the ringdown signals from binary black holes' merger to be detected with future GW detectors, for which we will focus on TianQin \cite{Luo:2015ght} and LISA \cite{Audley:2017drz}.

The paper is organized as follows.
In section \ref{sec:intro}, we introduce some basics of STVG, including the rotating black hole solution already known, then we obtain the corresponding QNMs.
In section \ref{sec:para.esti}, we study how the GR deviating parameter in STVG can be constrained with TianQin and LISA using the ringdown signal from the merger of binary black holes.
In section \ref{sec:sum}, we give a brief discussion and summary.

\section{Rotating black hole solution in STVG and its QNM}\label{sec:intro}

The action of STVG is given by \cite{Moffat:2005si}
\begin{equation}
S = \int d^4x(\mathcal{L}_G+\mathcal{L}_\phi+\mathcal{L}_S+\mathcal{L}_M),
\end{equation}
where $\mathcal{L}_M$ is the Lagrangian density of matter, and
\begin{eqnarray}
\mathcal{L}_{S}&=& \sqrt{-g} \bigg\{\frac{1}{G^3}\Big[\frac12g^{\mu\nu}\nabla_{\mu}G\nabla_{\nu}G-V(G)\Big]\nonumber\\
&&\qquad+\frac1{\mu^2 G}\Big[\frac{1}{2}g^{\mu\nu}\nabla_{\mu} \mu\nabla_{\nu}\mu -V(\mu) \Big]\bigg\}\,,\nonumber\\
\mathcal{L}_{G}&=&\frac{1}{16\pi G}\sqrt{-g} (R+2\Lambda)\,,\nonumber\\
\mathcal{L}_{\phi}&=&\frac{1}{4\pi} \sqrt{-g}  \Big[\frac{1}{4}B^{\mu\nu}B_{\mu\nu}-\frac{1}{2}\mu^2\phi^\mu\phi_\mu+V_\phi(\phi)\Big]\,,
\end{eqnarray}
are the Lagrangian densities of the scalar, tensor and vector fields, respectively.
The fields $G(x),\mu(x)$ are scalars related to Newton's constant $G_N$ and the mass of the vector field $\phi_\mu$, respectively, $B_{\mu\nu} =\partial_{\mu}\phi_{\nu} -\partial_{\nu}\phi_{\mu}$, and $V(G),V(\mu),V_\phi(\phi)$ are potentials.

A rotating black hole solution in the theory has been constructed for the special case $G=G_N(1+\alpha)$, $\mu\approx0$ and $V(G)=V(\mu)=V_\phi(\phi)=0\,$. In this case the action reduces to that of the Einstein-Maxwell theory,
\begin{equation}
S =\frac1{16\pi G}\int d^4x\sqrt{-g}\,\Big(R+2\Lambda+G B^{\mu\nu}B_{\mu\nu}\Big)\,,
\label{action}
\end{equation}
and the rotating black hole solution is nothing but the Kerr-Newman solution with a special choice of its charge parameter \cite{Moffat:2014aja}:
\begin{eqnarray}
ds^2&=&-\frac{\Delta_S}{\rho^2}(dt-a\, {\rm sin}^2 \theta d\phi)^2+\frac{\rho^2}{\Delta_S}dr^2\nonumber\\
&&+\frac{{\rm sin}^2\theta}{\rho^2}[(r^2+a^2)d\phi-a dt]^2+\rho^2 d\theta^2\,,\nonumber\\
\Delta_S&=&r^2-2G M r+a^2+GQ_\phi^2\,,\nonumber\\
\rho^2&=&r^2+a^2 {\rm cos}^2\theta\,,\label{solution}
\end{eqnarray}
where $a=J/M$ with $J$ being the angular momentum. The conserved charge of the vector field is assumed to be proportional to the mass \cite{Moffat:2014aja}
\begin{equation}
Q_\phi=\pm(\alpha G_N)^{1/2}M\,.\label{prop}
\end{equation}

The action (\ref{action}) differs from that of GR in two ways. Firstly, the vector field, $\phi_\mu$, in (\ref{action}) is to be distinguished from the usual electromagnetic field. Secondly, the coupling constant $G$ is different from Newton's gravitational constant $G_N$. In the case of $\alpha=0$, the vector field $\phi_\mu$ vanishes, $G$ returns to $G_N$ and the gravitational perturbation of (\ref{solution}) returns to that of a Kerr black hole in GR. By studying the QNMs of the ringdown signal, one can impose constraint on the GR deviating parameter $\alpha\,$.

The QNMs of the Kerr-Newman black hole have been studied with various methods, including a consideration of the static case \cite{Manfredi:2017xcv}. Dias et al. have studied the QNMs of the Kerr-Newman black hole with the Newton-Raphson method, where they solved the perturbation equations directly without variable separation \cite{Dias:2015wqa}. Dudley and Finley (DF) have obtained approximate decoupled equations for gravitational perturbations of the Kerr-Newman black hole \cite{Dudley:1978vd}. With these equations, Berti and Kokkotas have computed the QNMs with both the WKB and the continued fraction method \cite{Berti:2005eb,Kokkotas:1993ef}. The QNMs of the weakly charged \cite{Mark:2014aja} and the slow-rotation \cite{Pani:2013ija, Pani:2013wsa} solutions have also been calculated. The Newton-Raphson method is relatively more precise, but solving the DF equations with the continued fraction method is more efficient and we shall adopt this approach in the present work.

To test STVG, we need to find out the dependence of QNMs' complex frequencies $\Omega_{lm}$ on the parameter $\alpha$.  Following \cite{Berti:2005eb}, the DF equations of (\ref{solution}) are given by
\begin{eqnarray}
0&=&\frac{d}{du}\bigg[(1-u^2)\frac{dS_{lm}}{du}\bigg]+\bigg[a^2 \Omega_{lm}^2 u^2\nonumber\\
&&-2a\Omega_{lm} su+s+E_{lm}-\frac{m+su}{1-u^2}\bigg]S_{lm}\,,\nonumber\\
0&=&\Delta_S^{-s}\frac{d}{dr}\bigg[\Delta_S^{s+1}\frac{dR}{dr}\bigg]+\frac{1}{\Delta_S}\bigg[K^2-i s\frac{d\Delta_S}{dr}K\nonumber\\
&&+\Delta_S\bigg{(}2i s\frac{dK}{dr}-E_{lm}-a^2\Omega_{lm}^2 +2am\Omega_{lm}\bigg{)}\bigg]R\,,\label{DF}
\end{eqnarray}
where $u={\rm cos}\theta$, $K=(r^2+a^2)\Omega_{lm}-am$, and $E_{lm}$ is the separation constant. Note $s=-2$ corresponds to the gravitational perturbation.
Imposing the purely ingoing and outgoing boundary conditions at the black hole horizon and the space infinity, respectively, the angular function $S_{lm}$ and the radial function $R$ can be constructed as
\begin{eqnarray}
S_{lm}(u)&=&e^{a\Omega_{lm} u}(1+u)^{|m-s|/2}(1-u)^{|m+s|/2}\nonumber\\
&&\times\sum_{n=0}^\infty a_n(1+u)^n\,,\nonumber\\
R(r)&=&e^{i\Omega_{lm} r}(r-r_{-})^{-1-s+i\Omega_{lm}+i\sigma_+}(r-r_+)^{-s-i\sigma_+}\nonumber\\
&&\times \sum_{n=0}^{\infty}d_n\bigg{(}\frac{r-r_+}{r-r_-}\bigg{)}^n\,,\label{ansatz}
\end{eqnarray}
where $r_\pm=GM\pm(G^2M^2-a^2-GQ_\phi^2)^{1/2}$ are the two roots of $\Delta_S=0$, and $\sigma_+=[\Omega_{lm}(2GMr_+-GQ_\phi^2)-am]/(r_+-r_-)\,.$

Plugging (\ref{ansatz}) into (\ref{DF}), one can obtain two three-term continued fraction relations, which can be solved numerically for $\Omega_{lm}$ and $E_{lm}$ for each QNM. This method works more reliably for $Q_\phi\leq M/2$ \cite{Berti:2005eb}, corresponding to $\alpha\in[0,1/4]$. Frequencies for the fundamental modes with $l=m=2$ are listed in TABLE \ref{tab:QNMw22}, where we have set $G_N=M=1\,$. We can find that it is exactly the same result of Kerr-Newman if we replace $\alpha$ with $Q_\phi$ by \eqref{prop}. This is an obvious result since the metric and the equation are the same as Kerr-Newman with this replacement. Besides, if $\alpha=0$ we can recover the QNMs result of Kerr black hole.

\begin{table*}
\centering
\caption{Frequencies for the fundamental quasi-normal modes with $l=m=2$.}
\begin{tabular}{c c c c c c c}
\hline\hline
a&$\alpha$=0&$\alpha$=0.05&$\alpha$=0.1&$\alpha$=0.15&$\alpha$=0.2&$\alpha$=0.25\\
\hline
0& 0.373672 -0.088962&0.359461 -0.085110&0.346353 -0.081574&0.334221 -0.078316&0.322955 -0.075305&0.312465 -0.072513\\
0.2&0.402145 -0.088311&0.387457 -0.084463&0.373892 -0.080928&0.361322 -0.077670&0.349636 -0.074657&0.338741 -0.071863\\
0.4&0.439842 -0.086882&0.424807 -0.083012&0.410909 -0.079453&0.398019 -0.076169&0.386025 -0.073128&0.374834 -0.070303\\
0.6&0.494045 -0.083765&0.479241 -0.079765&0.465594 -0.076068&0.452972 -0.072637&0.441267 -0.069443&0.430384 -0.066459\\
0.8&0.586017 -0.075630&0.574636 -0.070848&0.564788 -0.066266&0.556428 -0.061827&0.549555 -0.057467&0.544233 -0.053118\\
0.96&0.767674 -0.049434&-&-&-&-&-\\
\hline
\end{tabular}
\label{tab:QNMw22}
\end{table*}

\begin{figure}
\centering
\includegraphics[width=0.45\textwidth]{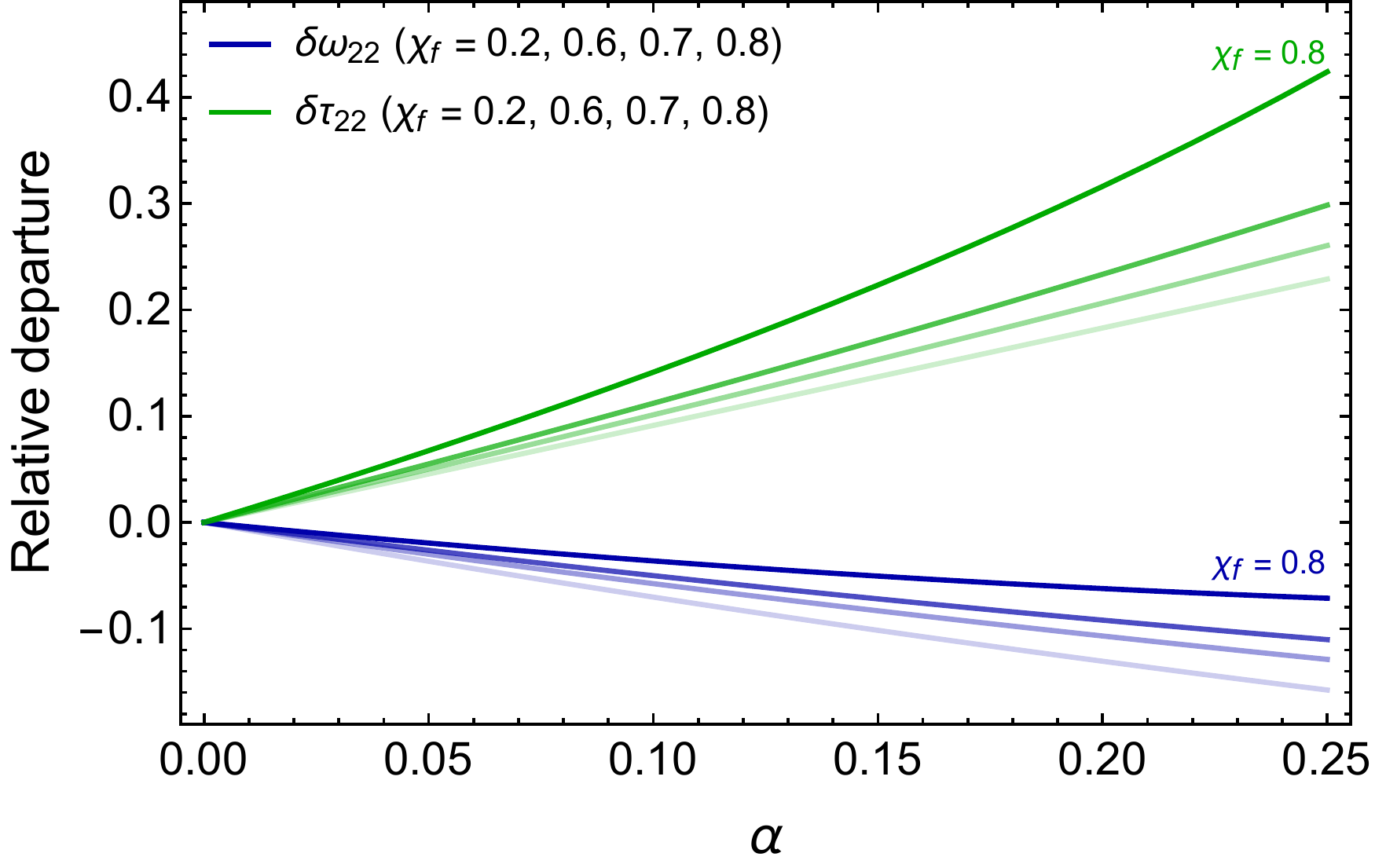}
\includegraphics[width=0.45\textwidth]{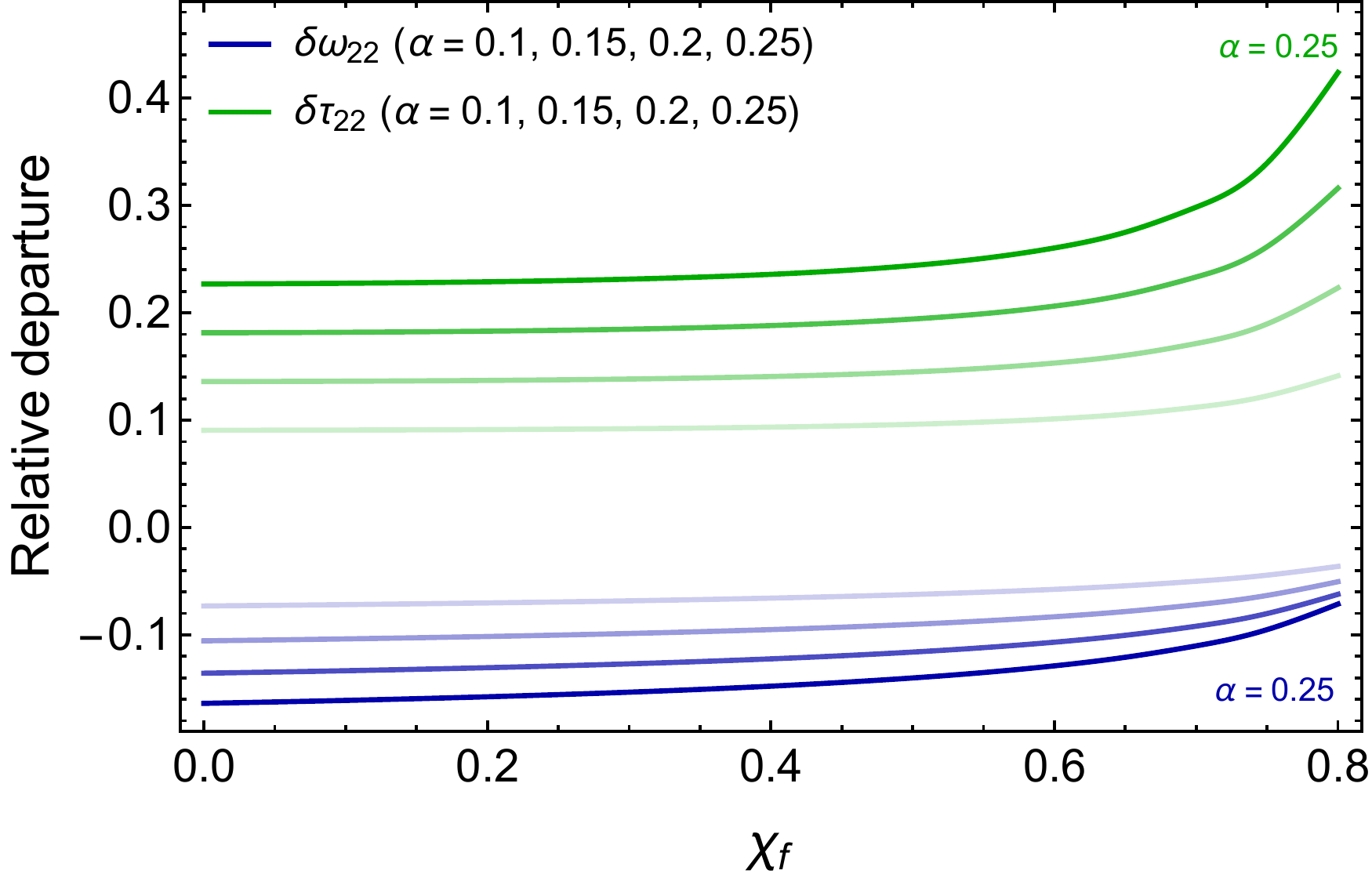}
\caption{Dependence of the relative departures $\delta w_{22}$ and $\delta\tau_{22}$ on $\alpha$(top) and $\chi_f$(bottom).
The lighter lines correspond to smaller values for the fixed parameter.}
\label{fig:deviation}
\end{figure}

An illustration of the dependence of the deviations of QNM frequencies for $l=m=2$ mode on the coupling constant $\alpha$ and final spin $\chi_f$ is given in FIG. \ref{fig:deviation}, where $\delta \omega_{lm}=[\omega_{lm}(\alpha)-\omega_{lm}(0)]/\omega_{lm}(0)$ and $\delta\tau_{lm}=[\tau_{lm}(\alpha) -\tau_{lm}(0)]/\tau_{lm}(0)\,$.
These are the parameters used in theory-agnostic parameterized framework which describe the frequency deviations on GR results.
Such framework has been discussed in several work, such as \cite{Shi:2019hqa, Gossan:2011ha, Meidam:2014jpa, Li:2011cg, Carullo:2018sfu, Brito:2018rfr}.
We can find that for special values of spin, the deviations of $\omega_{22}$ and $\tau_{22}$ are almost proportional to $\alpha$.
On the other hand, for small $\chi_f$ the relative departure is almost the same for fixed $\alpha$.
But for large $\chi_f$ such as $\chi_f>0.6$,
the deviation for $\tau_{22}$  will grow with $\chi_f$,
and the deviation for $\omega_{22}$  will decrease with $\chi_f$.
The deviations for other modes are similar to $22$ mode, thus we do not show them here.

For later convenience, we decompose the complex quasi-normal frequency into real and image part as
$\Omega_{lm}=\omega_{lm}+i/\tau_{lm}$,
and fit the numerical result with a set of phenomenological formulae similar to those in \cite{Berti:2005ys}, the
\begin{eqnarray}
M\omega_{lm}=f_1(1-\chi_f)^{f_2}+f_3+\alpha[f_4(1-\chi_f)^{f_5}+f_6]\,,\nonumber\\
Q_{lm}=q_1(1-\chi_f)^{q_2}+q_3+\alpha[q_4(1-\chi_f)^{q_5}+q_6]\,,\label{fitting.formulae}
\end{eqnarray}
where $Q_{lm}=\omega_{lm}\tau_{lm}/2$, and $\chi_f=J/(GM^2)$ is the dimensionless spin of the remnant black hole. The constants $f_i$ and $q_i$ are listed in TABLE \ref{tab:fit}, where the last column indicates the maximal percentage of error of the fit formulae from the true values.

\begin{table}[!htbp]
\centering
\caption{Constants and maximal error for (\ref{fitting.formulae}).}
\begin{tabular}{c c c c c c c c}
\hline\hline
$lm$&$f_1$&$f_2$&$f_3$&$f_4$&$f_5$&$f_6$&\%\\
\cline{1-8}
22&-1.2749&0.1197&1.6388&0.0615&5.2626&-0.2592&3.13\\		
21&-0.3194&0.2712&0.6856&0.0448&4.8229&-0.2554&1.99\\		
33&-1.3078&0.1894&1.8907&0.1244&4.1520&-0.4466&3.56\\
44&-2.7346&0.1109&3.5309&0.0698&6.4496&-0.5616&2.89\\
\hline\hline
$lm$&$q_1$&$q_2$&$q_3$&$q_4$&$q_5$&$q_6$&\%\\
\cline{1-8}
22&1.4422&-0.4941&0.6693&0.3268&-1.6119&-0.1761&1.64\\
21&1.6886&-0.2919&0.3941&-1340.6&0.0007&1340.7&8.97\\
33&2.3010&-0.4837&0.9527&0.5876&-1.5442&-0.4343&2.96\\
44&3.0104&-0.4910&1.3079&0.6208&-1.6737&-0.2787&1.54\\
\hline
\end{tabular}
\label{tab:fit}
\end{table}

\section{Constraining STVG with TianQin and LISA}\label{sec:para.esti}

In this section, we study how the GR deviating parameter $\alpha$ in STVG can be constrained using the ringdown signal from the merger of massive black holes to be detected with future space-based GW detectors, focusing on TianQin \cite{Luo:2015ght} and LISA \cite{Audley:2017drz}.

\subsection{Detectors}

The first detector we consider is TianQin \cite{Luo:2015ght}, which be a constellation of three satellites on a geocentric orbit with radius about $10^5$ kilometers. We adopt the following model for the sky averaged sensitivity of TianQin \cite{Luo:2015ght,Hu:2018yqb,Wang:2019ryf,Shi:2019hqa},
\begin{align}
\label{curveTianQin}
S_n(f)=\frac{10}3\Big[\frac{4S_a}{(2\pi f)^4L_0^2}\Big(1+\frac{10^{-4}{\rm Hz}}{f}\Big) +\frac{S_x(f)}{L_0^2}\Big]\nn\\
\times\Big[1+\Big(\frac{2fL_0/c}{0.41}\Big)^2\Big],
\end{align}
where \(L_0=\sqrt{3}\times 10^8{\rm m}\), $c$ is the speed of light, \(\sqrt{S_a}=1\times 10^{-15}{\rm ms^{-2}Hz^{-1/2}}\) is the average residual acceleration on each test mass and \(\sqrt{S_x}=1\times10^{-12}~{\rm mHz^{-1/2}}\) is the total noise of displacement measurement in a single laser link. To avoid the problem of having telescopes pointing to the Sun, TianQin adopts a ``3 month on + 3 month off'' observation scheme. To fill up the observation gap, one may consider having a twin set of TianQin constellations to operate consecutively. Such a scheme will not affect the sensitivity of each detector.

The second detector we consider is LISA \cite{Audley:2017drz}, the concept of which has been around for several decades. For LISA, we shall use the sensitivity curve given in \cite{Cornish:2018dyw}.

\subsection{Waveform of the ringdown signal}

The ringdown waveform is given by
\bea
h_+(t)&=&\frac{M_z}{D_L}\sum_{l,m>0}A_{lm}e^{-t/\tau_{lm}}\nn\\
&&\times Y_{+}^{lm}cos(\omega_{lm}t-m\phi)\,,\nn\\
h_\times(t)&=&-\frac{M_z}{D_L}\sum_{l,m>0}A_{lm}e^{-t/\tau_{lm}}\nn\\
&&\times Y_{\times}^{lm}sin(\omega_{lm}t-m\phi)\,,
\eea
where $M_z$ is the red-shifted mass of the remnant black hole,
$D_L$ is the luminosity distance to the source, $Y_{+,\times}^{lm}=Y_{+,\times}^{lm}(\iota)$ is the sum of spin -2 weighted spherical harmonics \cite{Kamaretsos:2011um}, with  $\iota\in[0,\pi)$ being the angle between the spin-axis of the source and the line-of-sight to the source, $\phi\in[0,2\pi] $ is the initial orbital phase of the source, and $A_{lm}$ is the amplitude of the $lm$ mode \cite{Meidam:2014jpa}
\bea
A_{22}(\nu)&=&0.864\nu\,,\nn\\
A_{21}(\nu)&=&0.43[\sqrt{1-4\nu}-\chi_{eff}]A_{22}(\nu)\,,\nn\\
A_{33}(\nu)&=&0.44(1-4\nu)^{0.45}A_{22}(\nu)\,,\nn\\
A_{44}(\nu)&=&[5.4(\nu-0.22)^2+0.04]A_{22}(\nu)\,,
\eea
where $\nu=m_1m_2/(m_1+m_2)^2$ is symmetric mass ratio, $$\chi_{eff}=\frac12\Big[\sqrt{1-4\nu}\chi_1 +\frac{m_1\chi_1-m_2\chi_2}{m_1+m_2}\Big]$$ is the effective spin, and $m_1,m_2,\chi_1,\chi_2$ are the masses and dimensionless aligned spins of the binary black hole before merger.
We use four leading modes to construct the ringdown signal, which are $22,21,33,44$ mode respectively. The chosen of modes follows \cite{Shi:2019hqa, Gossan:2011ha, Meidam:2014jpa, Carullo:2018sfu}, and more detailed discussions can be found therein.

\subsection{Statistical method}

The SNR for a GW signal is obtained with the following formula,
\bea
\label{SNR}
SNR[h]=\sqrt{(h|h)}\,,
\eea
and the inner product for any pair of signals \(p(f)\) and \(q(f)\) is defined as
\bea
\label{inner product}
(p|q)=2\int_{f_{\rm low}}^{f_{\rm high}} \frac{p^*(f)q(f)+p(f) q^*(f)}{S_n(f)}df\,,
\eea
where $f_{\rm low}$ is taken to be half the frequency of the \((2,1)\) mode and $f_{\rm high}$ is taken to be twice the frequency of the \((4,4)\) mode,  to prevent the ``junk'' radiation that occurs in the Fourier transformation \cite{Gossan:2011ha}.

In the case of large SNR, the uncertainty in parameter estimation is given by
\bea
\label{estimation}
\Delta\theta^a\equiv\sqrt{\langle \delta\theta^a \delta\theta^a\rangle}\approx\sqrt{(\Gamma^{-1})^{aa}}
\eea
where \(\theta^a\) are parameters to be estimated, \(\langle\dots\rangle\) denotes the expectation value, and $\Gamma^{-1}$ is the inverse of the  Fisher information matrix (FIM),
\bea
\label{FIM}
\Gamma_{ab}=\Big(\frac{\partial h}{\partial \theta^a}\Big|\frac{\partial h}{\partial \theta^b}\Big)\,.
\eea

We will focus on the sky-averaged result and the parameter space to be considered is
\begin{equation}
\overrightarrow{\theta}=\{M, \chi_f, D_L, \nu, \chi_{eff}, t_0, \phi, \iota, \alpha\},
\end{equation}
where $t_0$ is the time of coalescence.

It is also interesting to consider the combined constraint from all events that can be detected throughout the lifetime of a detector. Assuming that all the detected events are independent of each other, we can construct a combined FIM to study the cumulative constraint on $\alpha$. The parameter space is
\bea
\overrightarrow{\theta}&=&\{M_{z1}...M_{zi}\,;\chi_{f1}...\chi_{fi}\,;D_{L1}...D_{Li}\,;\nu_1...\nu_i\,;\nn\\
 &&\chi_{eff1}...\chi_{effi}\,;t_{01}...t_{0i}\,;\phi_1...\phi_i\,;\iota_1...\iota_i\,;\alpha\}\,.
\eea

\subsection{Results}

\begin{figure}[!htbp]
\begin{minipage}[c]{0.5\textwidth}
\includegraphics[width=0.7\textwidth]{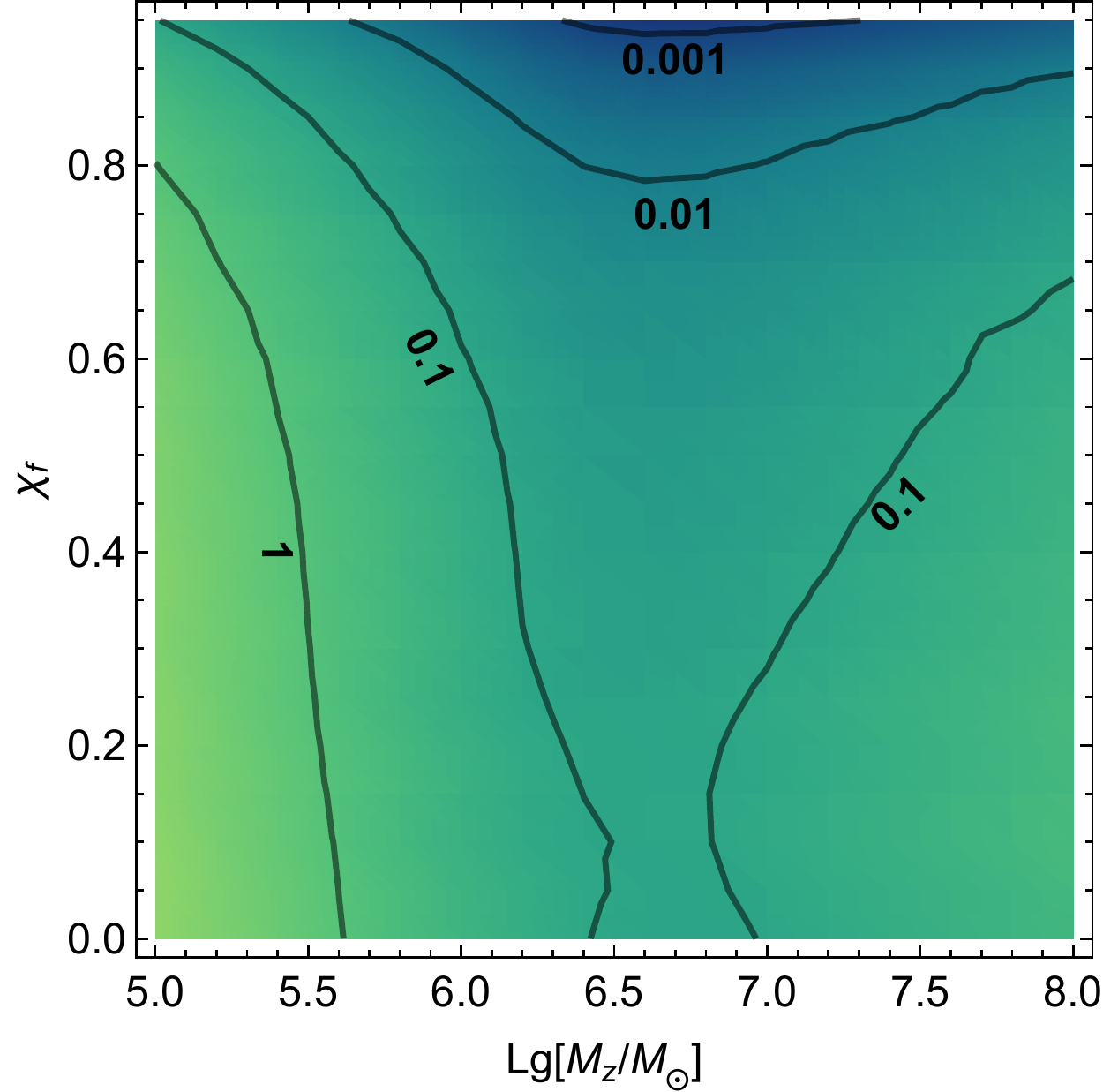}
\includegraphics[width=0.15\textwidth]{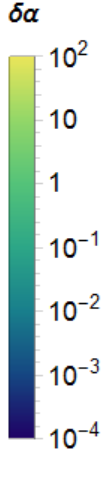}
\end{minipage}
\caption{The projected constraint of TianQin on $\delta\alpha$, varying with $\chi_f$ and $M_z$.}
\label{fig:contour1}
\end{figure}

\begin{figure}[!htbp]
\begin{minipage}[c]{0.5\textwidth}
\includegraphics[width=0.7\textwidth]{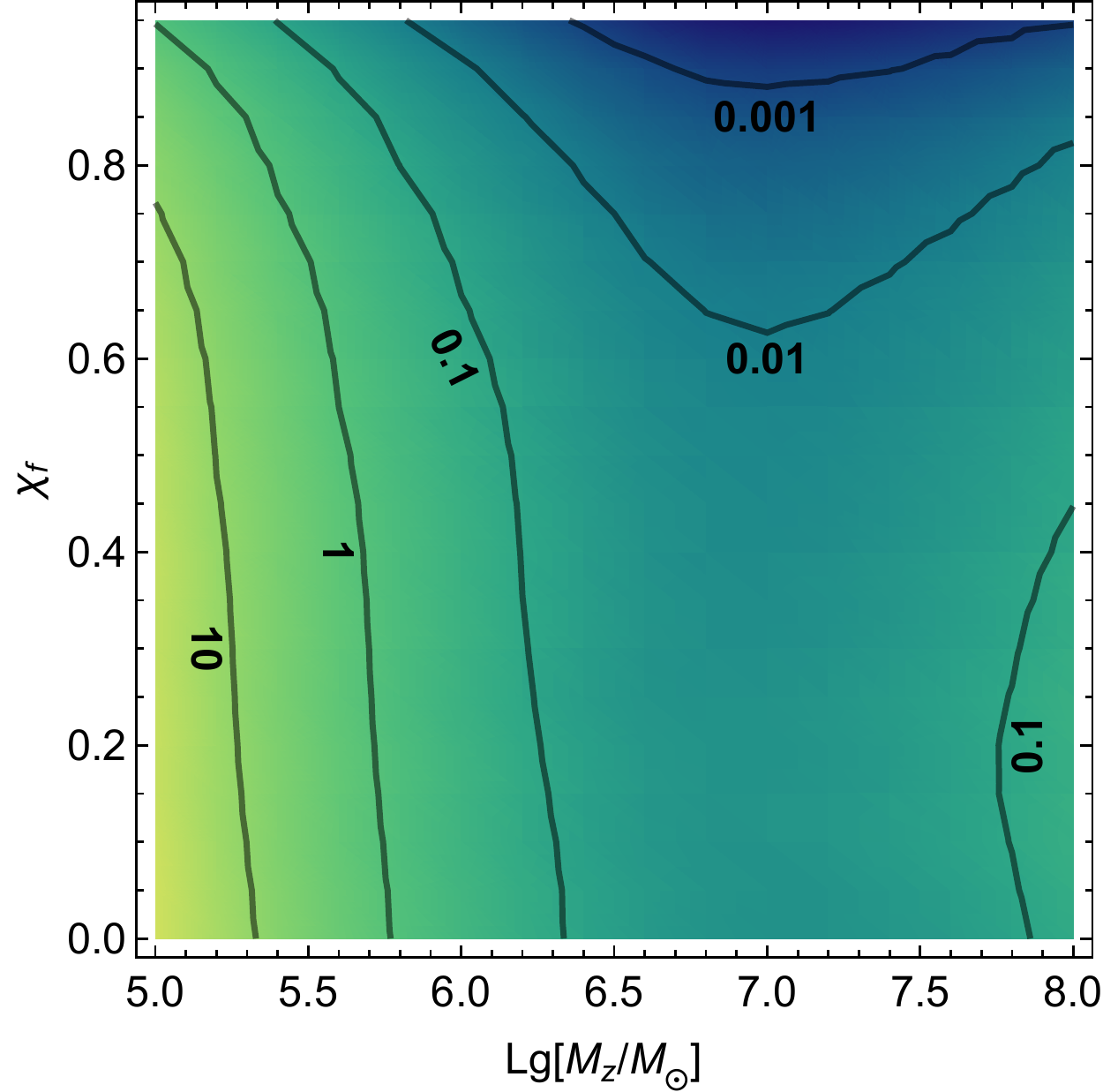}
\includegraphics[width=0.15\textwidth]{cb.png}
\end{minipage}
\caption{The projected constraint of LISA on $\delta\alpha$, varying with $\chi_f$ and $M_z$. }
\label{fig:contour2}
\end{figure}

The projected constraint of TianQin and LISA on the GR deviating parameter $\alpha$ with the detection of a single massive black hole merger is illustrated in FIG. \ref{fig:contour1} and FIG. \ref{fig:contour2}, respectively. In both figures, $\delta\alpha$ indicates the deviation of $\alpha$ from 0, and other parameters are: $D_L=15$Gpc, $\nu=2/9$, $\chi_{eff}=0.3$, $\iota=\pi/3$ and $t_0=\phi=0$. One can see that, $\alpha$ is best constrained with TianQin for $M_z\sim 4\times10^6M_{\odot}\,$ and with LISA for $M_z\sim10^7M_{\odot}\,$.

In optimal scenarios, both LISA and TianQin are expected to detect hundreds of massive black hole mergers throughout their mission lifetime \cite{Klein:2015hvg,Salcido:2016oor,Wang:2019ryf,Feng:2019wgq}. The projected number of massive black hole mergers that can be detected are largely model dependent. We use three models for the merger history of massive black holes as has been considered in \cite{Wang:2019ryf}. These models are denoted as ``popIII", ``Q3\_d" and ``Q3\_nod", corresponding to the light seed model \cite{Madau:2001sc} and the heavy seed models \cite{Bromm:2002hb,Begelman:2006db,Lodato:2006hw} with and without time delay between the merger of massive black holes and that of their host galaxies, respectively. Further explanation of these models can be found in \cite{Wang:2019ryf} and references therein.

We shall consider several different detector scenarios, including TianQin operating for a nominal lifetime of 5 years (``TQ"), a twin set of TianQin constellation operating for 5 years (``TQ\_tc"), LISA operating for 4 years (``LISA\_4y") and LISA operating for 10 years (``LISA\_10y").

For each of the detector scenarios, we produce 100 simulated catalogue from each of the models for the merger history of massive black holes. Each simulated catalogue is consisted of all the events that can be detected with the corresponding detector scenario (selected if the SNR of the whole waveform is greater than 8). Each data set gives a combined constraint on the GR deviating parameters $\alpha$. For a given detector scenario, one can average over the corresponding 100 sets of data to obtain an averaged constraint on $\alpha$. The results are listed in TABLE \ref{tab:combined}.

\begin{table}[!htbp]
\caption{The projected constraint on $\delta\alpha$ with different detector scenarios coupled with various models for the growth and merger history of massive black hole binaries. See text for further explanation.}\label{tab:tab2}
\centering
\begin{tabular}{c c c c}
\hline\hline
                         & pop\uppercase\expandafter{\romannumeral3}     & Q3\_nod     &Q3\_d\\
\hline
TQ        & $0.0331\pm0.0222$ & $0.0063\pm0.0027$ & $0.0101\pm0.0054$ \\
TQ\_tc    & $0.0181\pm0.0114$ & $0.0047\pm0.0017$ & $0.0066\pm0.0031$ \\
LISA\_4y  & $0.0066\pm0.0029$ & $0.0020\pm0.0008$ & $0.0041\pm0.0016$ \\
LISA\_10y & $0.0037\pm0.0021$ & $0.0011\pm0.0006$ & $0.0026\pm0.0001$ \\
\hline
\end{tabular}
\label{tab:combined}
\end{table}

\section{Summary and Discussion}\label{sec:sum}

To sum up, we have presented an explicit example of using the ringdown signal from a binary black hole merger to constrain a modified theory of gravity, i.e. STVG. We find that both TianQin and LISA have the potential to constrain $\alpha$ to the level of a few percent or better.

There is a caveat with the result obtained. Since both the action (\ref{action}) and the solution (\ref{solution}) are essentially the same as those of a charged rotating black hole, the effect of $\alpha$ in STVG is degenerate with that of the electric charge of a Kerr-Newman black hole in a usual Einstein-Maxwell system. However, the electric charges of astrophysical black holes tend to be quickly reduced due to the quantum Schwinger pair-production effect \cite{Gibbons:1975kk, hanni1982limits} and the vacuum breakdown mechanism \cite{Goldreich:1969sb, Ruderman:1975ju, Blandford:1977ds}. E. Barausse {\it et al.} \cite{Barausse:2014tra} have presented a theoretical upper bound on the charge-to-mass ratio of black holes, $Q/M\sim10^{-3}$, corresponding to $\alpha\sim 10^{-6}\,$. So if future space-based GW detectors were to consistently find $\alpha$ significantly greater than the order of $10^{-6}\,$, the result is more likely due to a genuine STVG effect rather than the electric charges of black holes.

\begin{acknowledgments}
We thank Peng-Cheng Li, Yi-Fan Wang and Viktor T. Toth for helpful discussion and correspondence. We also thank Alberto Sesana and Enrico Barausse for sharing their simulated catalogue of massive black holes. This work has been supported by the Natural Science Foundation of China (Grant Nos. 91636111, 11690022, 11703098, 11805286, 11475064).
\end{acknowledgments}

\bibliographystyle{apsrev4-1}
\bibliography{STVG}

\end{document}